\documentclass[a4paper,11pt]{article}
\usepackage{amsfonts, amssymb, amsthm}

\def\beq{\begin{equation}}
\def\eeq{\end{equation}}
\def\bea{\begin{eqnarray}}
\def\eea{\end{eqnarray}}
\def\nn{\nonumber}

\def\F{{\cal F}}
\def\opp{{\hat {\mathtt  p}}}
\def\opx{{\hat {\mathtt  x}}}

\def\hh{H}
\def\pp{T}
\def\bb{B}
\def\cc{G}
\def\aaa{A}

\def\>#1{{\bf #1}}

\newcommand{\sect}[1]{\setcounter{equation}{0}\section{#1}}
\newcommand{\subsect}[1]{\subsection{#1}}

\newtheorem{proposition}{Proposition}[section]

\begin{document}

\begin{center}
{\LARGE{\bf{ Twisted  Conformal Algebra  so(4,2)}}}
\end{center}

\bigskip\bigskip

\begin{center}
N. Aizawa$^\dagger$, F.J. Herranz$^\ddagger$,
J. Negro$^\star$ and M.A. del Olmo$^\star$

\end{center}

\begin{center}
$^\dagger${\sl Department of Applied Mathematics,
Osaka Women's University,\\ Sakai, 590-0035  Osaka, Japan}\\

{e-mail:   aizawa@appmath.osaka-wu.ac.jp}
 
\medskip

$^\ddagger${\sl Departamento de F\'{\i}sica,  Universidad de
Burgos, \\ E-09006  Burgos, Spain}\\

{e-mail:  fjherranz@ubu.es}
 
\medskip

$^\star${\sl Departamento de F\'{\i}sica Te\'orica,  
Universidad de Valladolid,
\\
E-47011  Valladolid, Spain}\\
 
{e-mail:  jnegro@fta.uva.es, olmo@fta.uva.es}
\end{center}

\vskip 1.5cm
\centerline{\today}
\vskip 1.5cm

\bigskip

\begin{abstract}
A new twisted deformation, $U_z(so(4,2))$, of the conformal algebra of the
$(3+1)$-dimensional Minkowskian spacetime  is presented. This
construction is  provided by a classical $r$-matrix spanned by   ten
Weyl--Poincar\'e generators, which generalizes
non-standard quantum deformations previously obtained for $so(2,2)$
and $so(3,2)$. However, by introducing a  conformal null-plane basis
it is found that the twist can  indeed be supported by  an
eight-dimensional carrier subalgebra. By construction the
Weyl--Poincar\'e subalgebra remains  as a   Hopf subalgebra after
deformation.  Non-relativistic limits of $U_z(so(4,2))$ are shown to
be well defined and they
  give rise to new twisted conformal  algebras of Galilean and
Carroll spacetimes.  Furthermore a difference-differential  massless
Klein-Gordon (or wave) equation  
  with twisted conformal symmetry is constructed through deformed
momenta and position operators. The deformation parameter is
interpreted as the lattice step on a uniform Minkowskian spacetime
lattice discretized along   two basic null-plane directions.
\end{abstract}

\vfill\eject

\sect{Introduction}\label{introduccion}

Quantum groups have been applied to obtain deformations of spacetime
symmetries in a search of new deformed relativisictic theories beyond 
the usual Lie symmetries, where  deformation parameters have been
interpreted as fundamental quantities in  the field  of
quantum gravity at the Planck scale (sse e.g.~\cite{Majid}).   Amongst
the quantum  kinematical algebras we can distinguish between standard
or quasitriangular type 
\cite{Lukierskia,Giller,Lukierskib,olmo93,
olmo94,olmo94a,olmo94b,kgalilei}  and non-standard or triangular
type \cite{beyond,nulla,nullb}. The non-standard deformations are
twisted quantum algebras \cite{drinfeld83}. For all of them,  
 deformed coproducts indicate   some broken symmetry related with some
kind of interaction or correlation of the elementary systems
\cite{olmo01a}.

In the last years a great activity  is being developed to
obtain explicitly non-standard deformations of classical Lie
algebras. Amongst the new recent twists that have been constructed, 
we can mention the so called extended twists
\cite{vladimir99},  peripheric twists \cite{olmo99a,olmo99b},
quantum Jordanian twists \cite{olmo01} or 
parabolic twists \cite{vladimir01}.  Another possibility is to obtain
a sequence of twists by composing  chains of twists
\cite{olmo99,olmo02}. In general, all of these twist deformations use
Cartan--Weyl bases that, although are well adapted for semisimple
Lie algebras, in our opinion they  present two main objections. By one
hand, they preclude a clear application of  contraction theories
which are completely necessary, for instance, if non-relativist limits
are searched. On the other, they do not provide, in principle, natural
physical interpretations of deformations parameters. These and other
reasons underly the constructions of non-standard quantum
deformations in other physical bases such as kinematical and
conformal ones, that try to establish deformations of relativistic
and non-relativistic spacetime  symmetries.  In  this context
we find   several  twisted deformations for
$so(2,2)$~\cite{beyond,herranz00a}, 
$so(3,2)$~\cite{Lukierskic,herranz97}, and very recently for 
$so(4,2)$~\cite{Lukierskid,herranz02}, as well as  for other real
forms together with their contractions.  Moreover,
in~\cite{aizawa01a,aizawa01b,aizawa02}  the twist introduced in
\cite{kulish00} is particularized to kinematical algebras. In
particular, in  \cite{aizawa01a,kulish00} it is proved that the twisted
deformation of $so(3,2)$ presented there is equivalent to that
introduced in
\cite{herranz97}.

The aim of this paper is  to construct a new twisted deformation of
the  real  Lie algebra $so(4,2)$ in a  conformal framework that
extends previous known results for lower dimensional cases as well as
to study physically meaningfull contractions and interpretations of
the deformation parameter.

The paper is organized  as follows. In
section~\ref{twistdeformation} we briefly review  the basics on
twisted deformations.  Section~\ref{conformalso42} is devoted to
describe the conformal Lie algebra $so(4,2)$  in order   to set up 
the notation used   throughout the paper as well as to fix two
suitable bases for our purposes: the usual conformal Minkowskian basis
and a `conformal null-plane basis' defined as a natural extension
of  the Poincar\'e one~\cite{leutwyler}; the  Weyl--Poincar\'e
 subalgebra ${\cal{WP}}\subset so(4,2)$  is identified within the
latter basis.

In section~\ref{twistdeformationso42} we  generalize the
classical $r$-matrices of the non-standard quantum deformations of
$so(2,2)$~\cite{beyond} and $so(3,2)$~\cite{herranz97} to the 
$so(4,2)$ case. The resulting classical $r$-matrix  is spanned by ten
Weyl--Poincar\'e generators (all but a rotation one), however,  by
writing  it in the conformal null plane basis we find that the carrier
subalgebra $\cal L$ is actually    an eight-dimensional (8D)  one.
Next we introduce the associated   conformal twisting
element  which is a factorizable extended twist so that   the Hopf
structure and universal $\cal R$-matrix  of the twisted
algebra $U_z(so(4,2))$ is then computed (the deformation
parameter $z$ is such that   ${\rm e}^z\equiv q$). We remark that, 
by construction, both the carrier and Weyl--Poincar\'e  subalgebras
remain as Hopf subalgebras after deformation in such a manner that
$U_z({\cal L})\subset U_z({\cal{WP}})\subset U_z(so(4,2))$.

A non-linear change of basis allows us to rewrite the Hopf structure
of $U_z({\cal{WP}})$ in terms of the   usual conformal basis in
section~\ref{twistdeformationso42conf}. As a first byproduct, the
twisting element and the universal $\cal R$-matrix of
$U_z(so(4,2))$ are expressed in this   usual   basis in a very
`compact' form. Another relevant consequence is that in this last
form, well defined  non-relativistic limits (contractions) to twisted
conformal Galilei and Carroll algebras are  applied in
section~\ref{contracciones}; the latter can alternatively be seen as
the  $(4+1)$D version of the $(3+1)$D null-plane quantum Poincar\'e
algebra obtained in  \cite{nulla,nullb}. For both cases we obtain the
contracted Hopf structures, twisting elements as well as universal ${\cal
R}$-matrices in an explicit and close form.

  A `twisted analogue' of the wave or massless Klein--Gordon
equation is deduced in section~\ref{discretequation} with twisted
conformal invariance by
introducing  some appropriate difference-differential  momenta and
position operators. Such a construction is
performed in the usual   Minkowskian spacetime coordinates as well as
in the null-plane ones. Both systems convey some kind of
discretization of   spacetime, although the
latter exhibits more clearly  a physical interpretation of the
deformation parameter $z$ as the lattice constant on a uniform
relativistic spacetime lattice discretized along {\em two}
distinguished null-plane directions. Furthermore, from a dynamical
point of view within the null-plane framework we find that $z$ may
also be interpreted as a fundamental constant related to a time
evolution parameter.

  Some conclusions and remarks close the paper where we   establish an
explicit relationship between our  carrier subalgebra and twisting
element with those  introduced by  Kulish and Lyakhovsky 
\cite{kulish00} in relation with $sl(4)$ by making profit from the
isomorphism as complex forms of $sl(4)\simeq so(6)$. We also comment on
the new $\kappa$-conformal classical $r$-matrices obtained 
by Lukierski {\em et al}  in \cite{Lukierskid}.

\sect{Twisted Hopf algebras}
\label{twistdeformation}

Let ${\cal A}(m,\Delta ,\epsilon,S)$ be a Hopf algebra with
multiplication $m\colon {\cal A}\otimes {\cal A}\to {\cal A}$,
coproduct $\Delta \colon {\cal A}\to {\cal A} \otimes {\cal A}$,
counit $\epsilon \colon {\cal A}\to C$ and antipode $S \colon
{\cal A}\to {\cal A}$. By using an invertible element (twisting)
${\cal F}\in {\cal A} \otimes {\cal A}$, such that ${\cal
F}=\sum  f_i^{(1)}\otimes f_i^{(2)}$, we can transform  the
original Hopf algebra into a new one, called {\em twisted Hopf
algebra}  \cite{drinfeld83}  ${\cal A}_{\cal F}(m,\Delta _{\cal
F},\epsilon ,S_{\cal F})$ endowed with the same multiplication
and counit,  but different coproduct and antipode given by:
\begin{equation}
\label{def-t}
 \Delta _{\cal F}(a)={\cal F}\Delta (a){\cal F}^{-1} \qquad
S_{\cal F}(a)= v S(a) v^{-1} \qquad v=\sum  f_i^{(1)}
S(f_i^{(2)})  \qquad\forall a \in {\cal A} .
\end{equation}
 The {\em twisting
element} ${\cal F}$ has to verify the following conditions
\begin{eqnarray}
\label{gentwist}
 (\epsilon \otimes  id)({\cal F}) = (id \otimes  \epsilon)({\cal
F})=1  \qquad
{\cal F}_{12}(\Delta \otimes  id)({\cal F}) =
{\cal F}_{23}(id \otimes  \Delta)({\cal F}) 
\end{eqnarray}
where ${\cal F}_{12}=\sum f_i^{(1)}\otimes f_i^{(2)}\otimes 1$,
${\cal F}_{23}=\sum 1\otimes f_i^{(1)}\otimes f_i^{(2)}$, and
so on. The first relation of (\ref{gentwist}) is a normalization
condition and follows from the second one modulo a non-zero
scale factor. Hence the $\cal R$-matrix associated to the
twisted algebra starting  either from a non-deformed Hopf
algebra  or from a deformed one (with universal $\cal R$-matrix
$\cal R$) is given, in this order, by
\beq\label{uniR}
{\cal R}_{\cal F}= \tau({\cal F}){\cal F}^{-1}  \qquad
{\cal R}_{\cal F}= \tau({\cal F}){\cal R} {\cal F}^{-1} 
\eeq 
where $\tau$ is the permutation of the tensor product factors.
So,  $\tau({\cal F})=\sum  f_i^{(2)}\otimes f_i^{(1)}$.

 It is worthy to note that if $\cal A$ is a Hopf subalgebra of a
Hopf algebra $\cal B$ the twist element $\cal F$ induces a
twisted deformation of $\cal B$. Let ${\cal A}={\cal U}({\cal
L}) \subset {\cal B}={\cal U}({\cal G})$ universal enveloping
algebras of Lie algebras ${\cal L}\subset {\cal G}$. When ${\cal
U}({\cal L})$ is the minimal subalgebra such that the twisting
element is completly defined as ${\cal F} \in {\cal U}({\cal L})
\otimes {\cal U}({\cal L})$ it is said that ${\cal L}$ is the
{\em carrier subalgebra} of ${\cal F}$
\cite{gerstenhaber93}.

Most of the known twisting elements are 
 {\em factorizable} with respect
to the coproduct. In this paper we will be interested in the
particular subclass of  factorizable twisting elements that 
verify  the following factorized twist equations \cite{RES}:
\begin{equation}\label{fact}
\begin{array}{l}
(\Delta \otimes id)(
{\cal F}) = {\cal F}_{13}{\cal F}_{23}\qquad
(id \otimes \Delta_{{\cal F}})({\cal F}) = {\cal F}_{12}{\cal
F}_{13} 
\end{array}
\end{equation}
which combined with the quantum Yang--Baxter equation on 
${\cal F}$  guarantee the validity of the twist equations
(\ref{gentwist}).

Two interesting cases of this kind of factorizable twists are
Jordanian and  extended Jordanian twists. A {\em Jordanian
twist}   with a 2D carrier algebra generated by $H$ and $E$ 
such that $[H,E]=E$ (Borel algebra) is characterized by the
canonical twisting element
\cite{ogiev93}
\begin{equation}\label{jordanian}
{\cal F}_{\cal J}= {\rm e}^{H \otimes \sigma}  \qquad
\sigma=\ln (1+E).
\end{equation}
The most known example   is related with
the
non-standard quantization of $sl(2)$ \cite{drinfeld87}.

 The problem of extending the  Jordanian twist was studied
in~\cite{vladimir99} by  considering a  carrier solvable
subalgebra ${\cal L}$ with, at least, four generators 
$\{H,E,A,B\}$, and non-vanishing commutation relations given by
\begin{equation}
\label{fourcarrier}
 [H,E]=2 E \qquad  [H,A]=\alpha A \qquad [H,B]=\beta B \qquad
[A,B]=\gamma E 
\qquad \alpha + \beta = 2.
\end{equation}
There exists an {\em extended Jordanian twist} for ${\cal
U}({\cal L})$ given by
\beq
\label{extended}
{\cal F}_{\cal E}= \Phi_{\cal J}  \Phi_{\cal E}=
{\exp}({H\otimes \sigma})\,
\exp({A\otimes B\, {\rm e}^{-2 \sigma}}).
\eeq

  New twisting elements can also be constructed by means of 
composition of  several twists, but note that in general the
composition of twists is not a twist.   However, given two
twists ${\cal F}_1$ and ${\cal F}_2$  with carrier algebras
 ${\cal L}_1$ and ${\cal L}_2$, respectively,  the composition  
 ${\cal F}_2\circ {\cal F}_1$ is a twist if the primitive
elements of    ${\cal L}_2$ are in the twisted
algebra   ${\cal U}_{{\cal F}_1}({\cal L}_1)$.

\sect{Conformal null-plane basis for so(4,2)}
\label{conformalso42}

The Lie algebra $so(4,2)$ can physically  be interpreted  either
as    the kinematical algebra of the  $(4+1)$D anti-de Sitter
spacetime or   as the conformal algebra of the  ($3+1$)D
Minkowskian  spacetime. We shall consider the conformal
interpretation and profit it  in order to use a `physical' basis
instead of the  Cartan--Weyl one, which   is the basis commonly
used in  the twist constructions.  

In the conformal basis we have the generators of rotations
$J_i$,   dilations $D$,    time translations $P_0$,   space
translations $P_i$,   boosts $K_i$ and    conformal
transformations $C_{\mu}$ (greek and latin labels take the
standard values: $\mu = 0, 1,2, 3$; $i =1, 2, 3$).  The 
non-vanishing commutation relations  of $so(4,2)$ read
\beq 
\label{so42}
\begin{array}{lll}
[J_i, J_j] =  \epsilon_{ijk} J_k &\qquad
  [J_i, K_j]  =  \epsilon_{ijk} K_k &\qquad
  [J_i, P_j]= \epsilon_{ijk} P_k  \\[0.2cm]
  [J_i, C_j] =  \epsilon_{ijk} C_k &\qquad
 [K_i, K_j] =- \epsilon_{ijk} J_k &\qquad
  [K_i, P_i]=   P_0    \\[0.2cm]
  [K_i, P_0] = P_i &\qquad
 [K_i, C_0] = C_i &\qquad
 [K_i, C_i] =  C_0 \\[0.2cm]
 [P_0, C_0] =  D &\qquad   [P_0, C_i]= -K_i &\qquad
 [P_i, C_0] = K_i \\[0.2cm]
 [P_i, C_j] = -\delta_{ij} D +  \epsilon_{ijk} J_k &\qquad 
 [P_{\mu}, D]= -P_{\mu} &\qquad
 [C_{\mu}, D]  = C_{\mu} 
\end{array}
\eeq
where hereafter sum over repeated indices should be  understood.

 As  is well known $so(4,2)$ contains a ($3+1$)D Poincar\'e
subalgebra ${\cal P}$ spanned by the ten generators
$\{J_i,P_\mu,K_i\}$. If we enlarge  ${\cal P}$ with the dilation
generator, we find the Weyl--Poincar\'e subalgebra 
 ${\cal WP}$, that is, the Lie algebra of the group of
similitudes of the Minkoswkian spacetime. Hence we
have the sequence of subalgebras
${\cal P}\subset {\cal WP}\subset so(4,2)$.

In the further  construction of the twisted quantum $so(4,2)$
algebra we shall deal with a basis closely related to the
null-plane basis of the Poincar\'e algebra
${\cal P}$. So, we   briefly describe the  structure of  
${\cal P}$ in relation with  the null-plane evolution scheme
\cite{leutwyler}. Consider   the Minkowskian spacetime with 
coordinates $x=(x^0,x^1,x^2,x^3)$ and  metric $g =(g_{\mu\nu})=
{\rm diag}\,(+,-,-,-)$. The initial state of a quantum
relativistic system can be defined on a null-plane plane
$\Pi_n^\tau$ determined by $n\cdot x=\tau$, where
$n$ is a light-like vector and $\tau$ a real constant. If we
take    the light-like vector  $n=(\frac 12,\frac 12,0,0)$, the
coordinate system naturally adapted to  $\Pi_n^\tau$ is given by:
\beq
\label{nullcoo}
\begin{array}{l}
x^- = n\cdot x =\frac 12(x^0-x^1)=\tau  \qquad
x^+=\frac 12(x^0+x^1)\qquad  x_T = (x^2, x^3).
\end{array}
\eeq
Therefore a point $x\in \Pi_n^\tau$  will be labelled by $(x^+,
x^2,x^3)$. The particular null-plane with $\tau=0$, $\Pi_n^{0}$,
is invariant under the action of the boosts  generated by $K_1$
since the {\em transverse} coordinates $x_T$ are unchanged and ${\rm
e}^{\chi K_1}$ maps $x^+$ on ${\rm e}^{\chi} x^+$. The
Poincar\'e generators   can be classified according to their
{\em goodness} $\gamma$ which is directly characterized by their
transformation with respect to $K_1$, i.e., 
\beq
\label{good}
[K_1,X]=\gamma X \qquad X\in {\cal P}.
\eeq
This property provides the {\em null-plane basis} of $\cal P$
defined by
\beq\label{goodness}
\begin{array}{ll}
\gamma =  +1  : &\quad P_+=\frac 12(P_0+P_1) \qquad E_2=\frac
12(K_2-J_3) \qquad E_3=\frac 12 (K_3+J_2) 
\\[0.2cm]
 \gamma = 0  :  & \quad K_1 \qquad J_1 \qquad P_2 \qquad  P_3
\\[0.2cm]
\gamma = -1 : & \quad  P_-=\frac 12(P_0-P_1) \qquad F_2=\frac
12(K_2+J_3) \qquad F_3=\frac 12 (K_3-J_2) 
\end{array}
\eeq
By taking into account (\ref{so42}), the non-vanishing
commutation relations  of ${\cal P}$ in this null-plane basis
turn out to be
\beq\label{basea}
\begin{array}{lll}
[K_1,P_+]=P_+ &\qquad [K_1,P_-]=-P_- &\qquad [P_+,F_a]=-\frac
12 P_a \\[0.2cm] [K_1,E_a]=E_a &\qquad [K_1,F_a]=-F_a &\qquad
[P_-,E_a]=-\frac 12 P_a \\[0.2cm] [J_1,P_a]=\epsilon_{1ab}P_b
&\qquad  [J_1,E_a]=\epsilon_{1ab}E_b &\qquad 
[J_1,F_a]=\epsilon_{1ab}F_b \\[0.2cm]
[E_a,P_a]=P_+ &\qquad  [F_a,P_a]=P_- &\qquad
[E_a,F_b]=\frac 12 \delta_{ab}K_1-\frac 12 \epsilon_{1ab}J_1  
\end{array}
\eeq
where from now on the two indices   $a,b=2,3$. Each set of
generators with the same goodness spans a Poincar\'e subgroup
$R_\gamma$ (note that $R_{\pm 1}$ are Abelian). The subgroup
with semidirect product structure $S_+=R_0\odot R_{+1}$ is the
stability group of $\Pi_n^{0}$, and the three remaining
generators, that close $R_{-1}$, move this  plane. In
particular, $P_-$ transforms $\Pi_n^{0}$ into
$\Pi_n^{\tau}$ so that $x^-=\tau$ can be considered as an
evolution parameter playing the role of a {\em time}. Recall  that
the   null-plane quantum Poincar\'e algebra \cite{nulla,nullb}
was constructed using a null-plane basis associated to $K_3$.

Now we extend the above  basis   of ${\cal P}$ to a
{\em conformal null-plane basis} of $so(4,2)$ by adding
$D,C_\mu$:
$$
{\cal P}=\{K_1,J_1,P_\pm,P_a,E_a,F_a\} \quad\to\quad so(4,2)=
\{\hh_\pm,J_1,\pp_\pm,\pp_a,\aaa_a,\bb_a,\cc_\pm,\cc_a\}  
$$
which in terms  of the initial conformal generators of $so(4,2)$
is defined by three sets of generators:
\beq\label{ccarrier}
\begin{array}{ll}
\hh_+=\frac 12 (D+K_1) &\qquad   \hh_-=\frac 12 (D-K_1)  \\[0.2cm]
\pp_+= \frac 12(P_0+P_1) & \qquad \pp_-= \frac 12(P_0-P_1)\\[0.2cm]
\pp_2=\frac 12 (P_2-P_3)&\qquad \pp_3=\frac 12 (P_2+P_3)\\[0.2cm]
\bb_2= \frac 12(K_2+J_3-K_3+J_2) &\qquad \bb_3= \frac
12(K_2+J_3+K_3-J_2) 
\end{array}
\eeq
\beq 
\begin{array}{l}
J_1 \qquad \aaa_2= \frac
12(K_2-J_3-K_3-J_2) \qquad  \aaa_3 =\frac
12(K_2-J_3+K_3+J_2)
\label{resto}
\end{array}
\eeq
\beq 
\begin{array}{ll}
\cc_+=\frac 12 (C_0+C_1)&\qquad 
\cc_-=\frac 12 (C_0-C_1)\\[0.2cm] \cc_2=\frac 12 (C_2-C_3)
&\qquad \cc_3=
\frac 12 (C_2+C_3) .
\label{restoc}
\end{array}
\eeq
The eight generators  (\ref{ccarrier}) span  the (carrier)
subalgebra $\cal L$ that we shall consider in the next section
as support of the twisting element of $so(4,2)$. These
generators  together with (\ref{resto}) give rise to the
Weyl--Poincar\'e subalgebra $\cal {WP}$. Finally, the new
conformal transformations (\ref{restoc}) complete the conformal
null-plane basis of $so(4,2)$.  Notice also  the following
relations with  the null-plane generators of 
${\cal P}$ (\ref{goodness}):
\beq
\pp_\pm=P_\pm  \qquad  
 \aaa_2=E_2-E_3 \qquad \aaa_3=E_2+E_3
 \qquad \bb_2=F_2-F_3  \qquad \bb_3=F_2+F_3 .
\eeq

 The goodness (\ref{good}) of these new generators reads
(cf.~(\ref{goodness})): 
\beq\label{confnull}
\begin{array}{ll}
 \gamma =  +1  : &\quad \pp_+  \quad \aaa_2 
 \quad \aaa_3 \quad \cc_+ \\[0.2cm]
  \gamma = 0  :  & \quad \hh_+  \quad 
 \hh_-\quad J_1\quad \pp_2\quad \pp_3\quad
\cc_2 \quad \cc_3 \\[0.2cm]
 \gamma = -1 : &\quad \pp_- \quad \bb_2
 \quad \bb_3 \quad \cc_-
\end{array}
\eeq

Starting from (\ref{so42}), we compute  the  commutations
relations   of $so(4,2)$ in terms of the conformal null-plane
basis (\ref{ccarrier})--(\ref{restoc}). The non-vanishing
commutators     are  displayed in three sets as follows: 

\noindent
$\bullet$ Carrier subalgebra ${\cal
L}=\{\hh_\pm,\pp_\pm,\pp_a,\bb_a\}$:
 \beq \label{carrier}
\begin{array}{llll}
 [\hh_+,\pp_+]=\pp_+ &\qquad   [\hh_+,\pp_a]=\frac 12 \pp_a
&\qquad   
[\hh_+,\bb_a]=-\frac 12\bb_a  &\qquad  [\bb_a,\pp_+]= \pp_a 
\\[0.2cm] [\hh_-,\pp_-]=\pp_- &\qquad   [\hh_-,\pp_a]=\frac 12
\pp_a &\qquad   [\hh_-,\bb_a]=\frac 12\bb_a  &\qquad 
[\bb_a,\pp_a]=  \pp_-  . 
\end{array}
\eeq

\noindent
$\bullet$ Commutations relations between  ${\cal L}$ and  
$\{\aaa_a,J_1\}$  (with the above ones (\ref{carrier}) close
$\cal {WP}$):
 \beq \label{weylp}
\begin{tabular}{lll}
$[\hh_+,\aaa_a]=\frac 12\aaa_a $ &\qquad 
$[\aaa_a,\pp_a]=\pp_+$ &\qquad
$[J_1,\aaa_a]=\epsilon_{1ab} \aaa_b$ \\[0.2cm]
$[\hh_-,\aaa_a]=-\frac 12\aaa_a $ &\qquad  
$[\aaa_a,\pp_-]= \pp_a$&\qquad  
$[J_1,\bb_a]=\epsilon_{1ab} \bb_b$ \\[0.2cm]
$[J_1,\pp_a]=\epsilon_{1ab} \pp_b$ &
\multicolumn{2}{l}{\qquad
 $[\aaa_a,\bb_b]=\delta_{ab}(\hh_+ - \hh_-)-
\epsilon_{1ab} J_1  .$} 
\end{tabular}
\eeq

\noindent
 $\bullet$ Commutators involving the new conformal
transformations $\{\cc_\pm ,
 \cc_a\}$:
\beq \label{conf}
\begin{tabular}{lll}
 $[\hh_+,\cc_-]= -\cc_- $ &\qquad  $[\hh_+,\cc_a]=-\frac 12\cc_a
$  &\qquad   $[\cc_+,\pp_-]=-\hh_-$ \\[0.2cm]
 $[\hh_-,\cc_+]= -\cc_+ $ &\qquad  $[\hh_-,\cc_a]=-\frac 12\cc_a
$  &\qquad   $[\cc_-,\pp_+]=-\hh_+$ \\[0.2cm]
 $[\cc_+,\pp_a]= -\frac 12 \aaa_a $ &\qquad 
$[\cc_+,\bb_a]=-\cc_a $  &\qquad   $[\cc_a,\pp_+]=\frac 12
\aaa_a$ \\[0.2cm]
 $[\cc_-,\pp_a]= -\frac 12 \bb_a $ &\qquad 
$[\cc_-,\aaa_a]=-\cc_a $  &\qquad   $[\cc_a,\pp_-]=\frac 12
\bb_a$ \\[0.2cm]
$[\cc_a,\aaa_a]= -\cc_+ $ &\qquad  $[\cc_a,\bb_a]=-\cc_- $ 
&\qquad $[J_1,\cc_a]=\epsilon_{1ab} \cc_b$  \\[0.2cm]
 \multicolumn{3}{l}{$[\cc_a,\pp_b]=\frac 12 \delta_{ab}(\hh_+ +
\hh_ -)+
\frac 12 \epsilon_{1ab} J_1   .$   } 
\end{tabular}
\eeq

 Similarly to   the null-plane Poincar\'e basis,  generators
with the same goodness close a $SO(4,2)$ subgroup  
$M_\gamma$;   $M_{\pm 1}$ are again Abelian.

\sect{Twisting deformation of  so(4,2)  in a conformal
null-plane basis}
\label{twistdeformationso42}

Non-standard quantum deformations for lower dimensional
conformal algebras of the Minkowskian  spacetime have already
been introduced by starting  from the Jordanian classical
$r$-matrix of the Borel algebra \cite{ogiev93}. The results
cover the $(1+1)$D case with   $so(2,2)$ \cite{beyond}, the
 $(2+1)$D case with $so(3,2)$ \cite{herranz97} and also the
corresponding quantum conformal algebras for the Galilean
spacetime obtained by means of contraction  (or non-relativistic
limit). The   non-standard (triangular) classical
$r$-matrices underlying these quantum algebras are given by
 \beq
\begin{array}{ll}
U_z({\cal WP}(1+1))\subset U_z(so(2,2)):&\quad
r=z(D\wedge P_0 + K_1\wedge P_1) \\[0.2cm]
U_z({\cal WP}(2+1))\subset U_z(so(3,2)):&\quad
r=z(D\wedge P_0+K_1\wedge P_1 + K_2\wedge P_2) + z J\wedge P_2  
\end{array}
\label{ab}
\eeq
where $z$ is the deformation parameter. By construction, the
Weyl--Poincar\'e subalgebra remains as   a Hopf subalgebra after
deformation since  the generators involved in the $r$-matrices
span  ${\cal WP}$. In fact, the presence of the dilation
generator $D$ in these $r$-matrices is essential in order to
fulfil the classical Yang--Baxter equation. This is worthy to be
compared with the standard (quasitriangular) quantum
deformations of the Poincar\'e algebra, such as the
$\kappa$-Poincar\'e \cite{Lukierskia,Giller,Lukierskib}, that
do not require the presence of $D$.  This kind of properties 
have been  studied for  other  Drinfeld--Jimbo deformations  in
\cite{Vlado}. Recall also that other different non-standard
classical $r$-matrices for $so(3,2)$ and $so(4,2)$ in a conformal
framework have been obtained in \cite{Lukierskic},  and  very recently 
a detailed study of classical $r$-matrices giving rise to 
$\kappa$-deformations of $so(4,2)$ with $\kappa$-Weyl--Poincar\'e
Hopf subalgebras  has been presented in \cite{Lukierskid}.

The generalization of the classical $r$-matrices (\ref{ab}) to
the  $(3+1)$D case reads
\beq 
r=z(D\wedge P_0+K_1\wedge P_1 + K_2\wedge P_2 + K_3\wedge
P_3)+z(J_3\wedge P_2 - J_2 \wedge P_3)  
\label{ac}
\eeq
which is again formed by  generators  of
${\cal WP}$.  In order to deduce the corresponding twisted Hopf
algebra of $so(4,2)$, we shall adopt the conformal null-plane
basis (\ref{ccarrier})--(\ref{restoc}) associated to
$K_1$, for which the
$r$-matrix becomes
\beq
 r=2 z (\hh_+\wedge \pp_+ + \hh_-  \wedge \pp_- + \bb_2\wedge
\pp_2  + \bb_3\wedge \pp_3) .
\label{aac}
\eeq
 Hence we have a 8D carrier subalgebra $\cal L$ with generators
(\ref{ccarrier}) and non-vanishing Lie brackets (\ref{carrier})
instead of the 11D subalgebra ${\cal WP}$.

 It is worthy to remark that two other algebraically equivalent 
 classical $r$-matrices can be obtained starting from
(\ref{ac}): keep the first term and  change the second one by
another with permuted indices as  $J_1\wedge P_3 - J_3 \wedge
P_1$ or $J_2\wedge P_1 - J_1 \wedge P_2$. These possibilities
would be related to null-plane basis associated to
$K_2$ or $K_3$, respectively.

 We proceed now to obtain the coproduct for 
the Hopf algebra $U_z(so(4,2))$ in three steps: first for the
generators of the   carrier Hopf subalgebra  $U_z({\cal L})$ by
means of the twisting element; second   for   $\{\aaa_a,J_1\}$
that complete the  Hopf subalgebra $U_z({\cal {WP}})$; and
finally for the  remaining conformal transformations. Once the
Hopf structure of $U_z(so(4,2))$ is found, we shall change back
to the initial conformal basis, where contractions  and a study
of difference-differential equations can meaningfully be 
performed.

\subsect{Twisting element and carrier Hopf subalgebra}
\label{8dtwist}

The twisting element associated to the carrier subalgebra 
 ${\cal L}=\{\hh_\pm,\pp_\pm,\pp_a,\bb_a\}$, with underlying
classical  $r$-matrix (\ref{aac}), is composed by three factors
\beq 
\label{twist}
 \F = \Phi_{\cal A}\, \Phi_{\cal E}\, \Phi_{\cal J}  
\eeq
where  
\bea 
\label{twistfactor}
 & & \Phi_{\cal J} = {\rm e}^{\hh_- \otimes \sigma_-}    \nn
\\[0.1cm]
 & & \Phi_{\cal E}\; = \exp(-2 z  \bb_3 \otimes \pp_3 \,
{\rm e}^{-\frac{1}{2}\sigma_-})
 \exp(-2 z  \bb_2 \otimes \pp_2 \,{\rm
e}^{-\frac{1}{2}\sigma_-}) \\[0.2cm] 
& & \Phi_{\cal A}\; = {\rm
e}^{\hh_+ \otimes \sigma_+}. \nn
\eea
with 
\beq \label{twistfactor1}
\sigma_- = \ln(1-2 z \pp_-)  \qquad 
\sigma_+ = \ln(1-2 z \pp)  \qquad
 \pp = \pp_+ +z(\pp_2^2+\pp_3^2){\rm e}^{-\sigma_-}. 
\eeq
The term $\Phi_{\cal J} $ is a Jordanian twist
(\ref{jordanian}),  $\F_{\cal E} \equiv \Phi_{\cal E} \Phi_{\cal
J}$ is an extended twist (\ref{extended}) and  $\Phi_{\cal A}$
is an additional Jordanian twist on the deformed carrier  space
of the extended twist. The explicit structure of $\F$ is the
generalization   of the $(2+1)$D case associated to $so(3,2)$
introduced in~\cite{aizawa01a,aizawa01b,kulish00}.

 After progresive application of the three twist factors on the
primitive (non-deformed) coproducts $\Delta (X)=X\otimes
1+1\otimes X $,  we obtain  deformed coproducts  $\Delta_{\cal
F}$ for the generators of ${\cal L}$.  Useful  relations for  
computations are given by
\[
\begin{tabular}{lll}
$[\hh_+,\sigma_-]=0 $&\quad $[\hh_+, \pp] = \pp$ &\quad $[\hh_+, 
\sigma_+] = 1-{\rm e}^{-\sigma_+}$ \\[0.2cm]
$[\hh_-, \sigma_-] = 1-{\rm  e}^{-\sigma_-}$&\quad
$[\hh_-,\pp]= z(\pp_2^2+\pp_3^2)\,{\rm e}^{-2\sigma_-}$
&\quad $[\hh_-, \sigma_+] =- 2 z^2(\pp_2^2+\pp_3^2)\, {\rm  
e}^{-(\sigma_+ + 2\sigma_-)}$ \\[0.2cm]
$[\bb_a, \sigma_-] =0$&\quad  
$[\bb_a, \pp] =  \pp_a   \,  {\rm e}^{-\sigma_-} $ &\quad    
$[\bb_a, \sigma_+]
=-2 z\, \pp_a\,{\rm  e}^{-(\sigma_+ + \sigma_-)}$.
\end{tabular}
\]
Note also that the
 Abelian commutation relations for the translations generators
ensure that all the elements
$\{\pp_\pm,\pp_a,\pp,\sigma_+,\sigma_-\}$ commute amongst
themselves. The results are as follows:
\medskip

\noindent
$\bullet$ {\em Jordanian twist}  $\Delta_{\cal J} \equiv
\Phi_{\cal J} \Delta
\Phi_{\cal J}^{-1}$:
\beq \label{JT}
\begin{array}{ll}
\Delta_{\cal J}(\hh_+) =\hh_+ \otimes 1 + 1 \otimes \hh_+ 
&\qquad
 \Delta_{\cal J}(\pp_+) =  \pp_+ \otimes 1 + 1 \otimes \pp_+  
\\[0.2cm]
\Delta_{\cal J}(\hh_-) = \hh_- \otimes {\rm e}^{-\sigma_-} + 1
\otimes \hh_-  &\qquad
\Delta_{\cal J}(\pp_-) = \pp_- \otimes {\rm e}^{\sigma_-} + 1
\otimes \pp_-
\\[0.2cm]
\Delta_{\cal J}(\bb_a)= \bb_a \otimes {\rm e}^{\frac 12 \sigma_-
} + 1 \otimes \bb_a&\qquad
\Delta_{\cal J}(\pp_a) = \pp_a \otimes {\rm e}^{\frac
12\sigma_-} + 1 \otimes \pp_a .
\end{array}
\eeq
 
\noindent  
$\bullet$ {\em Extended twist}  $\Delta_{\cal E} = \Phi_{\cal E}   
\Delta_{\cal J}\Phi_{\cal E}^{-1}$:
\beq  \label{ET}
 \begin{array}{l}
\Delta_{\cal E}(\hh_+) = \hh_+ \otimes 1 + 1 \otimes \hh_+
\\[0.2cm]
 \Delta_{\cal E}(\hh_-) = \hh_- \otimes {\rm e}^{-\sigma_-} + 1
\otimes \hh_-
+ 2 z  \bb_2 \otimes \pp_2 \,{\rm e}^{-\frac 32\sigma_-}  + 2 z
\bb_3 \otimes \pp_3 \,{\rm e}^{-\frac 3 2 \sigma_-}\\[0.2cm]
\Delta_{\cal E}(\bb_a) =  \bb_a \otimes {\rm e}^{-\frac 12
\sigma_-} + 1 \otimes \bb_a   \\[0.2cm]
\Delta_{\cal E}(\pp_+)= \pp_+ \otimes 1 + 1 \otimes \pp_+ - 2 z
\pp_2 \otimes \pp_2\,{\rm e}^{-\frac 12 \sigma_-}- 2 z \pp_3
\otimes \pp_3 \,{\rm e}^{-\frac 12 \sigma_-}  \\[0.1cm]
 \qquad\qquad\qquad  +2 z^2 \pp_- \otimes
(\pp_2^2+\pp_3^2)\,{\rm e}^{-\sigma_-}  \\[0.2cm]
\Delta_{\cal E}(\pp_-) =  \pp_- \otimes {\rm
e}^{\sigma_-} + 1 \otimes \pp_-
\\[0.2cm]
\Delta_{\cal E}(\pp_a) =  \pp_a\otimes {\rm e}^{\frac 12
\sigma_-} + {\rm e}^{\sigma_-}
\otimes \pp_a.
\end{array}
\eeq
 
\noindent 
$\bullet$ {\em  Additional Jordanian twist}  $\Delta_{\F} = \Phi_{\cal A}
\Delta_{\cal E} \Phi_{\cal A}^{-1}$:
\beq 
\label{AJT}
\begin{array}{l}
\Delta_{\F}(\hh_+) = \hh_+ \otimes {\rm e}^{-\sigma_+} + 1 \otimes \hh_+ 
\\[0.2cm]
\Delta_{\F}(\hh_-) = \hh_- \otimes {\rm e}^{-\sigma_-} + 1 \otimes \hh_-
  + 2 z \bb_2 \otimes \pp_2\, {\rm e}^{-\frac
12 (\sigma_+ +3\sigma_-)}     + 2 z \bb_3 \otimes \pp_3\, 
{\rm e}^{-\frac 12 (\sigma_+ +3\sigma_-)}\\[0.2cm]
 \qquad\qquad\qquad  + 2 z^2 \hh_+  \otimes (\pp_2^2+\pp_3^2)\, {\rm
e}^{-(\sigma_+ +2\sigma_-)}  
\\[0.2cm]
\Delta_{\F}(\bb_a) = \bb_a \otimes {\rm e}^{-\frac 12 (\sigma_+
+\sigma_-)} + 1 \otimes \bb_a +2 z \hh_+\otimes  \pp_a\, {\rm
e}^{-(\sigma_+ +\sigma_-)} 
\\[0.2cm]
\Delta_{\F}(\pp_+) = \pp_+ \otimes {\rm e}^{\sigma_+} + 1 \otimes \pp_+
- 2 z 
\pp_2 \otimes \pp_2\,{\rm e}^{\frac 12 (\sigma_+ -\sigma_-)} - 2 z 
\pp_3 \otimes \pp_3\, {\rm e}^{\frac 12 (\sigma_+ -\sigma_-)}   \\[0.2cm]
  \qquad\qquad\qquad + 2 z^2 \pp_- \otimes (\pp_2^2+\pp_3^2)\,{\rm
e}^{-\sigma_-} 
\\[0.2cm]
  \Delta_{\F}(\pp_-) = \pp_- \otimes {\rm e}^{\sigma_-} + 1 \otimes \pp_-
\\[0.2cm]
  \Delta_{\F}(\pp_a) = \pp_a \otimes {\rm e}^{\frac 12 (\sigma_+
+\sigma_-)} + {\rm e}^{\sigma_-} \otimes \pp_a  .
\end{array} 
\eeq

The three coproducts for $\pp_-$, which are the same, show that 
$\sigma_-$ always remains  primitive 
$$
\Delta_{\cal J}(\sigma_-)=\Delta_{\cal E}(\sigma_-)
=\Delta_{\F}(\sigma_-) =\sigma_-\otimes 1 + 1\otimes \sigma_- . 
$$ 
The  element $\sigma_+$ is also primitive:
$\Delta_{\F}(\sigma_+)  =\sigma_+\otimes 1
+ 1\otimes \sigma_+ $, provided that
\beq
 \Delta_{\F}(\pp) = \pp \otimes {\rm e}^{\sigma_+} + 1
\otimes \pp  .
\label{twistE}
\eeq
 
Therefore, the final  twisted coproducts (\ref{AJT})  
together with commutation relations (\ref{carrier}) define the
twisted carrier  algebra $U_z({\cal L})$. Furthermore these
results allow us to state: 

\begin{proposition}
The  element $\F$ given by (\ref{twist})
is a factorizable twist that fulfils equations (\ref{fact}).
\end{proposition}

In~\cite{vladimir99} it was already proved that  the product
$\Phi_{\cal E}\,\Phi_{\cal J}$ satisfies such equations. Hence,
it is only necessary to analyze the third factor $\Phi_{\cal
A}$. It can be checked that it satisfies
\beq
(\Delta_{\cal E} \otimes id)(\Phi_{\cal A}) = (\Phi_{\cal
A})_{13}(\Phi_{\cal A})_{23} 
\qquad
  (id \otimes \Delta_{\F})(\Phi_{\cal A}) = (\Phi_{\cal A})_{12}
(\Phi_{\cal A})_{13}         
\label{factb}
\eeq
which guarantees the  twisting condition
\beq
 (\Phi_{\cal A})_{12} (\Delta_{\cal E} \otimes id)(\Phi_{\cal
A}) = (\Phi_{\cal A})_{23}(id
\otimes
\Delta_{\cal E})(\Phi_{\cal A}).
 \label{cond1}
\eeq
In other words,   $\Phi_{\cal A}$ is a Jordanian twisting element
for the deformed algebra obtained by an extended twist based on
the carrier subalgebra spanned by $\{ \hh_-, \pp_-, \pp_a,
\bb_a \}$ since $[\hh_+, \pp] = \pp$.

\subsect{Weyl--Poincar\'e Hopf subalgebra}
\label{wptwist}

Let us compute the coproducts for $J_1$ and $\aaa_a$. From
(\ref{weylp}) we firstly find some useful relations:
\[
\begin{array}{lll}
 &[J_1, \sigma_-]= [J_1, \pp]   = [J_1, \sigma_+] = 0
\qquad  &[\aaa_a,\sigma_-] = -2 z  \,\pp_a\, {\rm e}^{-\sigma_-}  
\nonumber\\[0.2cm]
&[\aaa_a,\pp] =  \pp_a (1-{\rm e}^{\sigma_+} ) {\rm
e}^{-\sigma_-}  \qquad
& [\aaa_a,\sigma_+] =  2 z   \,\pp_a (1-{\rm e}^{-\sigma_+} ) {\rm
e}^{-\sigma_-}  .
\end{array}
\]
Hence $J_1$ remains primitive for the  three twists:
\beq\label{coproJ1}
\Delta_{\cal J}(J_1) = \Delta_{\cal E}(J_1) = \Delta_{\F}(J_1) =
 J_1 \otimes 1 + 1 \otimes J_1.
\eeq
The three steps for the coproduct of $\aaa_a$ are given by
\bea
&& \Delta_{\cal J}(\aaa_a) = \aaa_a \otimes {\rm e}^{-\frac
12\sigma_-} + 1\otimes \aaa_a+  2 z \,\hh_- \otimes \pp_a 
\,{\rm e}^{-\sigma_-} \nn\\[0.2cm]
&& \Delta_{\cal E}(\aaa_a)\; =  \aaa_a
\otimes {\rm e}^{-\frac 12 \sigma_-} + 1 \otimes \aaa_a
+2 z \hh_+ \otimes \pp_a\, {\rm e}^{-\sigma_-}\nn\\[0.1cm]
&&\qquad\qquad\qquad  
+2 z \bb_a \otimes \pp\,{\rm e}^{-\frac 12 \sigma_-}  - 2 z
\epsilon_{1ab} J_1 \otimes \pp_b\, {\rm e}^{-\sigma_-}  
\nn\\[0.2cm]
&& \Delta_{\F}(\aaa_a)=  \aaa_a \otimes {\rm e}^{\frac
12(\sigma_+ -\sigma_-)} + 1 \otimes \aaa_a + 2 z \hh_+ \otimes
\pp_a \, {\rm e}^{-(\sigma_+ + \sigma_-)}
\nn\\[0.1cm] 
&&\qquad\qquad\qquad  
+2 z \bb_a \otimes \pp\,{\rm e}^{-\frac 12 (\sigma_+
+\sigma_-)} -2 z \epsilon_{1ab} J_1 \otimes   \pp_b\,
 {\rm e}^{-\sigma_-} . \label{addJTF}
\eea
Consequently,  the previous quantum carrier algebra
 $U_z({\cal L})$, determined by coproducts (\ref{AJT}) and
commutators  (\ref{carrier}),   enlarged with coproducts
(\ref{coproJ1}) and (\ref{addJTF})  together with the 
commutation rules (\ref{weylp}) determine, by construction, a
new  twisted Weyl--Poincar\'e algebra in a null-plane basis,
 $U_z({\cal WP})$,  such that $U_z({\cal L})  \subset U_z({\cal
WP})$.

\subsect{Conformal transformations}
\label{ctrans}

In the same way, cumbersome computations would lead to the
coproducts of the conformal transformations completing
$U_z(so(4,2))$. We restrict ourselves to display the auxiliary
commutation relations between the generators
$\{\cc_\pm,\cc_a\}$ and the elements  $\{\sigma_\pm,\pp\}$:
$$
\label{gsigma}
\!\!\!
\begin{array}{l}
[\cc_+,\sigma_-]=2z\,  {\rm e}^{-\sigma_-}\hh_- + 2 z^2\, {\rm
e}^{-2\sigma_-}\pp_-\qquad   [\cc_-, \sigma_-]=0\qquad
 [\cc_a,\sigma_-]=-z\,{\rm e}^{-\sigma_-}\bb_a\nonumber\\[0.2cm]
\displaystyle{ [\cc_+,\pp]=-z\,  {\rm
e}^{-\sigma_-}\left\{ \pp_2\aaa_2+\pp_3\aaa_3+\pp_+   + 2(\pp -
\pp_+ ) (\hh_- +    {\rm e}^{-\sigma_-})
\right\}  }\nonumber\\[0.2cm]
[\cc_-,\pp]=-\hh_+-z\,{\rm
e}^{-\sigma_-}(\pp_2\bb_2+\pp_3\bb_3+\pp_-)\nonumber\\[0.2cm]
 [\cc_a,\pp]=\frac 12 \aaa_a +
\displaystyle{  z\, {\rm
e}^{-\sigma_-}\left\{ \pp_a(\hh_+ + \hh_-) + \epsilon_{1ab}\pp_b
J_1 +  \pp_a({\rm e}^{-\sigma_-}-1) +(\pp-\pp_+)\bb_a  
\right\}}\nonumber\\[0.2cm]
\displaystyle{ [\cc_+,\sigma_+]=2 z^2\, 
 {\rm e}^{-(\sigma_+ +\sigma_-
)}\left\{ \pp_2\aaa_2+\pp_3\aaa_3+\pp_+   + 2(\pp - \pp_+ ) (\hh_-
+    {\rm e}^{-\sigma_-})
\right\}  \nonumber  }\\[0.2cm]
\qquad\qquad\qquad 
\displaystyle{ + 4z^3\, {\rm e}^{-(2\sigma_+ + \sigma_-)}
\left\{ {\rm e}^{-\sigma_-}(\pp - \pp_+)^2 +\pp (\pp - \pp_+)
\right\} }\nonumber\\[0.2cm]
\displaystyle{  [\cc_-,\sigma_+]=2z\, {\rm e}^{-\sigma_+}
\left\{ \hh_+ +
z\,{\rm e}^{-\sigma_-}(\pp_2\bb_2+\pp_3\bb_3+\pp_-) \right\}+
2z^2\, {\rm e}^{-2 \sigma_+} \left\{\pp+  {\rm e}^{-\sigma_-}
(\pp-\pp_+)  \right\}\nonumber }\\[0.2cm]
\displaystyle{  [\cc_a,\sigma_+]=-  z \,  {\rm
e}^{- \sigma_+  }  \aaa_a -4z^3\, {\rm e}^{-(2\sigma_+ +
\sigma_-)} \pp_a \left\{ \pp  + {\rm e}^{-\sigma_-} (\pp-\pp_+)
\right\} }\nonumber\\[0.2cm]
\qquad\qquad\qquad 
\displaystyle{  - 2z^2\, {\rm e}^{-(\sigma_+ + \sigma_-)}
\left\{\pp_a(\hh_+ +
\hh_-) +
\epsilon_{1ab}\pp_b J_1  +   \pp_a({\rm e}^{-\sigma_-}-1)
+(\pp-\pp_+)\bb_a  
\right\} }.
\end{array}
$$

 Finally, we remark that the universal quantum $\cal R$-matrix
for a twisted algebra is given in terms of the twisting
element by (\ref{uniR}). Hence, in our particular  case it reads
\beq \label{uniRso}
{\cal R}= \tau(\Phi_{\cal A})\, \tau( \Phi_{\cal
E})\,\tau(\Phi_{\cal J})\,
\Phi_{\cal J}^{-1}\, \Phi_{\cal E}^{-1}\, \Phi_{\cal A}^{-1} .
\eeq
The first order term of the  power series of $\cal R$ in $z$
reproduces the classical $r$-matrix (\ref{aac}). By construction
the three Hopf algebras in the sequence 
$U_z({\cal L})\subset  U_z({\cal {WP}}) \subset U_z(so(4,2))$ 
share the same  universal $\cal R$-matrix and classical
$r$-matrix.

\sect{ Weyl-Poincar\'e Hopf subalgebra  in the
standard conformal basis}
\label{twistdeformationso42conf}

The twisted Weyl--Poincar\'e algebra  $U_z({\cal WP})$, 
previously obtained in a null-plane
basis, can   be written in the initial conformal basis
(\ref{so42}) through the following non-linear map
\beq\label{wa}
\begin{array}{ll}
P_0=-\frac {1}{2 z}(\sigma_+ + \sigma_-)&\qquad 
P_1=-\frac {1}{2 z}(\sigma_+ - \sigma_-) \\[0.2cm]
P_2=(\pp_3+\pp_2) {\rm e}^{-\sigma_-}&\qquad
P_3=(\pp_3-\pp_2) {\rm e}^{-\sigma_-}\\[0.2cm]
D=\hh_+ +\hh_- &\qquad K_1=\hh_+ - \hh_-  \\[0.2cm]
K_2=\frac 12(\aaa_3+\aaa_2 +\bb_3+ \bb_2)&\qquad
K_3=\frac 12(\aaa_3-\aaa_2 +\bb_3-\bb_2)\\[0.2cm]
J_2=\frac 12(\aaa_3-\aaa_2 -\bb_3+ \bb_2)&\qquad
J_3=-\frac 12(\aaa_3+\aaa_2 -\bb_3- \bb_2)  
\end{array}
 \eeq
with $J_1$  unchanged. The limit $z\to 0$ gives rise to the 
change of basis (\ref{ccarrier}) and (\ref{resto}). This map is
the natural generalization of that introduced in
\cite{aizawa01a,aizawa01b} for $U_z({\cal WP}(2+1))$ in relation
to the non-standard quantum $so(3,2)$. 

 By taking into account the inverse map for the commuting
elements $\{\pp_\pm,\pp_a,\pp,\sigma_\pm\}$,
 \beq\label{wb}
\begin{array}{l}
\sigma_-=-z(P_0-P_1)\qquad 
\displaystyle{ \pp_-=\frac{1-  {\rm e}^{- zP_0}\,{\rm e}^{
zP_1}}{2z}}\qquad
\pp_2=\frac 12 \,  {\rm e}^{\sigma_-} (P_2 -P_3)\\[0.3cm]
\sigma_+=-z(P_0+P_1)\qquad \displaystyle{ \pp=\frac{1-  {\rm
e}^{- zP_0}\,{\rm e}^{- zP_1}}{2z} }\qquad \pp_3=\frac 12 \, 
{\rm e}^{\sigma_-} (P_2 +P_3)\\[0.3cm]
\displaystyle{  \pp_+=\frac{1-  {\rm e}^{- zP_0}\,{\rm e}^{
-zP_1}}{2z}-\frac z2
\,{\rm e}^{-zP_0}\,{\rm e}^{zP_1}(P_2^2+P_3^2) } ,
\end{array}
\eeq
the results of the previous section and the  auxiliary
commutators given by
\[
\begin{array}{lll}
[J_a, \sigma_-]=- z\epsilon_{1ab}P_b&\qquad
[J_a, \pp]=- \frac 12  \epsilon_{1ab}\,{\rm e}^{\sigma_+}P_b
&\qquad [J_a,\sigma_+]=z\epsilon_{1ab}P_b\\[0.2cm]
[K_a, \sigma_-]=- z P_a&\qquad
[K_a,\pp]= (1-\frac 12\, {\rm e}^{\sigma_+})P_a 
&\qquad [K_a,\sigma_+]=z(1-2\, {\rm e}^{-\sigma_+})P_a 
\end{array}
 \]
it can be found that  the coproduct and non-vanishing
commutation rules  for
$U_z({\cal WP})$ in the usual conformal basis turn out to be:
\beq\label{we}
\begin{array}{l}
\Delta(P_0)= P_0\otimes 1+1\otimes P_0\qquad
\Delta(P_1)= P_1\otimes 1+1\otimes P_1  \\[3mm]
\Delta(P_a)= P_a\otimes {\rm e}^{-z P_1}+1\otimes P_a  \qquad
\Delta(J_1)= J_1\otimes 1+1\otimes J_1  \\[3mm]
\Delta(J_a)= J_a\otimes {\rm e}^{-zP_1}+1\otimes J_a  +z J_1
\otimes P_a\\[3mm]
\Delta(D)=  D\otimes {\rm e}^{z P_0}\cosh zP_1+1\otimes D   +
K_1\otimes {\rm e}^{zP_0}\sinh zP_1+z(K_2+J_3)\otimes {\rm
e}^{zP_0} P_2 \\[1mm]
\qquad\qquad\qquad  +z(K_3-J_2)\otimes {\rm e}^{zP_0}P_3 +\frac
{z^2}2(D+K_1)\otimes  {\rm e}^{zP_0}\, {\rm
e}^{zP_1}(P_2^2+P_3^2)  \\[3mm]
\Delta(K_1)= K_1\otimes {\rm e}^{zP_0}\cosh zP_1 + 1\otimes K_1   +
D\otimes {\rm e}^{zP_0}\sinh zP_1- z(K_2+J_3)\otimes {\rm
e}^{zP_0} P_2\\[1mm]
\qquad\qquad\qquad  -z(K_3-J_2)\otimes {\rm e}^{zP_0}P_3 -\frac
{z^2}2(D+K_1)\otimes  {\rm e}^{zP_0} \,{\rm
e}^{zP_1}(P_2^2+P_3^2) \\[3mm]
\Delta(K_a)=  K_a\otimes {\rm e}^{zP_0} + 1\otimes K_a  +
\epsilon_{1ab}J_b\otimes ({\rm e}^{zP_0}-{\rm e}^{-zP_1}) 
- z \epsilon_{1ab}J_1\otimes P_b\\[1mm]
\qquad\qquad\qquad   + z(D+K_1)\otimes {\rm e}^{zP_0} \, {\rm
e}^{zP_1}P_a  
\end{array}
\eeq

\beq
\label{wf}
\begin{array}{l}
[J_i,J_j]=\epsilon_{ijk} J_k  \qquad
 [J_i, K_j]  =  \epsilon_{ijk} K_k \qquad
[K_i, K_j] =- \epsilon_{ijk} J_k\qquad [J_1, P_j]=
\epsilon_{1jk} P_k\\[3mm]
\displaystyle{[J_a, P_1]= \epsilon_{a1b} P_b\qquad
[J_a,P_b]=\epsilon_{1ab}\left(\frac{1-{\rm e}^{-2 z P_1}}{2
z}+\frac
z2\,(P_b^2-P_a^2)\right)+z\,\delta_{ab}(\epsilon_{1bc}P_bP_c)
  }\\[0.4cm]
[K_1,P_0]=\frac 1z \,{\rm e}^{zP_0}\sinh zP_1 -\frac z2\, {\rm
e}^{zP_0}\, {\rm e}^{zP_1}(P_2^2+P_3^2)  \\[3mm]
[K_1,P_1]=\frac 1z ({\rm e}^{zP_0}\cosh zP_1 -1)-\frac z2\, {\rm
e}^{zP_0}\,{\rm e}^{zP_1}(P_2^2+P_3^2)  \\[3mm]
[K_a,P_0]={\rm e}^{zP_0}\,{\rm e}^{zP_1}P_a\qquad
[K_a,P_1]=({\rm e}^{zP_0}\,{\rm e}^{zP_1}-1)P_a\qquad
[K_1,P_a]=(1-{\rm e}^{zP_0}\,{\rm e}^{-zP_1})P_a\\[3mm]
[K_a,P_b]=\delta_{ab}\left\{
\frac 1z \,{\rm e}^{-zP_1}({\rm e}^{zP_0}-\cosh
zP_1)+\frac z2 \,\epsilon^2_{1bc}(P_b^2-P_c^2)\right\}+z 
\epsilon^2_{1ab}P_aP_b  \\[3mm]
 [D,P_0]=\frac 1z ({\rm e}^{zP_0}\cosh zP_1
-1)+\frac z2\, {\rm e}^{zP_0}\,{\rm e}^{zP_1}(P_2^2+P_3^2) 
\\[3mm] 
[D,P_1]=\frac 1z \,{\rm e}^{zP_0}\sinh zP_1 +\frac z2\,
{\rm e}^{zP_0}\,{\rm e}^{zP_1}(P_2^2+P_3^2)\qquad 
[D,P_a]={\rm e}^{zP_0}\,{\rm e}^{-zP_1}\,P_a  ,
\end{array}
\eeq
(recall that $a,b,c=2,3$ while $i,j,k=1,2,3$). In this way we
find the $(3+1)$D version of the non-standard quantum ${\cal
WP}$  algebra obtained in
\cite{herranz97}.

Furthermore, the map (\ref{wa}) allows us
 to write the  universal ${\cal R}$-matrix of both $U_z({\cal
WP})$ and $U_z(so(4,2))$   (\ref{uniRso})  in this standard
conformal basis provided that the   three factors   
(\ref{twistfactor}) of ${\cal F}$ read
\beq\label{wg}\begin{array}{l}
\Phi_{\cal J}=\exp(-\frac z2\, D\otimes P_0)\exp(\frac z2\,
D\otimes P_1)
\exp(\frac z2\, K_1\otimes P_0)\exp(-\frac z2\, K_1\otimes P_1)\\[2mm]
\Phi_{\cal E}=\exp\left\{- z (K_2+J_3 )\otimes {\rm
e}^{-\frac{z}{2} P_0 }\,{\rm e}^{\frac{z}{2}P_1 } P_2 \right\} 
\exp\left\{-  z ( K_3-J_2)\otimes {\rm
e}^{-\frac{z}{2} P_0 }\,{\rm e}^{\frac{z}{2} P_1 } P_3 \right\}\\[2mm]
\Phi_{\cal A}= \exp(-\frac z2\, D\otimes P_0)\exp(-\frac z2\,
D\otimes P_1)
\exp(-\frac z2\, K_1\otimes P_0)\exp(-\frac z2\, K_1\otimes P_1) .
\end{array}\eeq
In the computations we have used the following expression
valid whenever
$[A,B]=0$:
\[
\exp\{(A+B)\otimes C\}=\exp\{ A\otimes C\} \exp\{ B\otimes C\}.
\]
Moreover,  the twisting element ${\cal F}$ (\ref{twist}) of
$U_z(so(4,2))$  can be obtained in a compact form by applying
the expression
\[
\exp(\alpha A\otimes B)\exp(\beta C\otimes D)=\exp\left\{
\beta C\otimes D {\rm e}^{\alpha\gamma B}\right\}
\exp(\alpha A\otimes B)  
\qquad \alpha,\beta,\gamma\in {\mathbb C}
\]
 valid whenever $[A,C]=\gamma C$ and $[B,D]=0$.
 The resulting ${\cal F}$  is given by
\beq\label{wi}
\begin{array}{l}
{\cal F}=\exp\left\{-  z (K_3-J_2)\otimes  {\rm e}^{z P_1 }
P_3\right\}
\exp\left\{- z (K_2+J_3 )\otimes 
{\rm e}^{z P_1 } P_2 \right\} \\[0.20cm] 
 \qquad\quad \times\exp(-z K_1\otimes P_1) \exp(-  z   D\otimes
P_0) 
\end{array}\eeq
which gives rise to the ${\cal R}$-matrix
${\cal R}=\tau({\cal F}){\cal F}^{-1}$ 
with   underlying classical $r$-matrix   given by (\ref{ac}).

\sect{Quantum contractions}
\label{contracciones}

The twisted Minkowskian  conformal algebra has well defined
non-relativistic limits to quantum conformal algebras of either
Galilean or Carroll spacetimes \cite{bacry}, denoted
$\cal G$ and $\cal C$, respectively. Such quantum contractions
are defined by the usual In\"on\"u--Wigner transformations  of
generators,  providing the Lie algebra contractions
$so(4,2)\to {\cal G}$ and $so(4,2)\to {\cal C}$,  together
with  a mapping on the deformation parameter $z$ that ensures
the convergency of the classical $r$-matrix
\cite{herranz97,gromov}. We stress that  computations can be
done once  $U_z({\cal WP})\subset U_z(so(4,2))$ are expressed in
the standard conformal basis presented in the previous section.


\subsect{Twisted conformal Galilean algebra $U_z({\cal G})$}

The  quantum {\em speed-space}  contraction that starting from
$U_z (so(4,2))$ leads to $U_z({\cal G})$ is defined by
\beq\label{na}
\begin{array}{llll}
 J_i\to J_i&\qquad P_0\to P_0&\qquad C_0\to C_0&\qquad D\to
D\\[2mm] P_i\to \varepsilon P_i&\qquad K_i\to \varepsilon
K_i&\qquad C_i\to \varepsilon C_i&\qquad z\to \varepsilon^{-2}z
\end{array}
\eeq
where $\varepsilon$ is the contraction parameter which is related
to the speed of light $c$ through $\varepsilon=1/c$.

Once the map (\ref{na}) is applied to {\em any}  element
associated to   $U_z (so(4,2))$, the limit $\varepsilon\to 0$ (or
$c\to \infty$)  gives rise to that corresponding to $U_z({\cal
G})$. Explicitly, the contraction of the classical $r$-matrix  
(\ref{ac}) originates the $r$-matrix
\beq\label{nc}
r=z(K_1\wedge P_1 + K_2\wedge P_2 + K_3\wedge 
P_3) .
\eeq
The coproduct and non-vanishing
commutation relations of $U_z({\cal G})$ turn out to be
\beq\label{nd}
\begin{array}{l}
\Delta(X)=X\otimes 1 + 1\otimes X\qquad
X\in\{J_i,K_i,P_\mu,C_i\}\\[2mm]
\Delta(D)=D\otimes 1 + 1 \otimes D + z(K_1\otimes P_1 +
K_2\otimes P_2 + K_3\otimes P_3)\\[2mm]
\Delta(C_0)=C_0\otimes 1 + 1 \otimes C_0 - z(C_1\otimes P_1 +
C_2\otimes P_2 + C_3\otimes P_3) 
\end{array}
\eeq
 \beq \label{ne}
\begin{tabular}{lll}
$[J_i, J_j] =  \epsilon_{ijk} J_k$ &\qquad  $ [J_i, K_j] = 
\epsilon_{ijk} K_k 
$&\qquad$ [J_i, P_j] = \epsilon_{ijk} P_k  $ \\[0.2cm]
$[J_i, C_j] =  \epsilon_{ijk} C_k $ &\qquad  $[K_i, P_0] = P_i
$&\qquad   $[K_i, C_0] = C_i$ \\[0.2cm]
$[P_0, C_0] = D$   &\qquad  $ [P_0, C_i] = -K_i$ &\qquad 
$[P_i, C_0] = K_i $\\[0.2cm]
$[P_i, D] = -P_i$  & 
\multicolumn{2}{l}{\qquad $[P_0, D] = -P_0-\frac
z2\,\>P^2$}\\[0.2cm] 
$[C_i, D] =  C_i$  & 
\multicolumn{2}{l}{\qquad $[C_0, D] =C_0-\frac
z2\,\>K^2  $  } 
\end{tabular}
\eeq
where  $\>P^2=P_1^2+P_2^2+P_3^2$ and
$\>K^2=K_1^2+K_2^2+K_3^2$.  Now commutation relations involving
conformal transformations are rather simplified. 

Under contraction the $\Phi$-factors (\ref{wg}) and twisting
element  ${\cal F}$  (\ref{wi}) reduce to
\beq\label{nf} 
\begin{array}{l}
\Phi_{\cal J}=\Phi_{\cal A}= \exp(-\frac z2\, K_1\otimes P_1)\\[2mm]
\Phi_{\cal E}=\exp (- z K_2 \otimes
 P_2 )\exp (- z K_3 \otimes P_3)   \\[2mm]
{\cal F}= \exp (- z K_3 \otimes P_3) \exp (- z K_2 \otimes P_2
)\exp(-z K_1\otimes P_1)
\end{array}
\eeq
which, in turn, determine  the universal $\cal R$-matrix of
$U_z({\cal G})$, which is simply the exponential of the classical
$r$-matrix (\ref{nc}) 
\beq\label{ng} {\cal R}  =  {\rm e}^{r}=\exp (z K_1 \wedge P_1)
\exp ( z K_2 \wedge P_2) \exp (z K_3 \wedge P_3) .
\eeq
Notice that the carrier subalgebra  of $U_z({\cal G})$   is  a
6D Abelian subalgebra  generated by $\{P_i, K_i\}$ and that the
Weyl--Galilean subalgebra ${\cal {WG}}$ spanned by
$\{J_i,P_\mu,K_i\}$ remains as a Hopf subalgebra,
$U_z({\cal {WG}})$, of $U_z({\cal {G}})$.


\subsect{Twisted conformal Carroll  algebra $U_z({\cal C})$}

The second non-relativistic limit $U_z (so(4,2)) \to U_z({\cal
C})$  corresponds to the quantum {\em speed-time} contraction 
   defined by the map
\beq\label{ma}
\begin{array}{llll}
J_i\to J_i&\qquad P_i\to P_i&\qquad C_i\to C_i&\qquad D\to D\\[2mm]
 P_0\to \varepsilon P_0&\qquad K_i\to \varepsilon K_i&\qquad
C_0\to \varepsilon C_0&\qquad z\to \varepsilon^{-1}z .
\end{array}
\eeq
The   classical $r$-matrix   for $U_z({\cal C})$ 
 reads
\beq\label{mb}
r=z(D\wedge P_0+K_1\wedge P_1 + K_2\wedge P_2 + K_3\wedge
P_3) .
\eeq
 The coproduct and non-vanishing commutation relations of
$U_z({\cal C})$ are given by
\beq\label{mc}
\begin{array}{l}
\Delta(X)=X\otimes 1 + 1\otimes X\qquad X\in\{J_i,P_\mu\}\\[2mm]
\Delta(Y)=Y\otimes {\rm e}^{z P_0} + 1\otimes Y\qquad
Y\in\{K_i,C_0\}\\[2mm]
\Delta(D)=D\otimes {\rm e}^{zP_0}+1\otimes D+  z K_1\otimes
{\rm e}^{zP_0} P_1  + z K_2\otimes {\rm e}^{zP_0} P_2 + z
K_3\otimes {\rm e}^{zP_0} P_3 \\[2mm]
\Delta(C_i)=C_i\otimes {\rm e}^{zP_0} + 1 \otimes C_i - z
C_0\otimes {\rm e}^{zP_0} P_i +z\,\epsilon_{ijk} K_j\otimes {\rm
e}^{zP_0}J_k
\end{array}
\eeq
\beq \label{md}
\begin{tabular}{lll}
$ [J_i, J_j] =  \epsilon_{ijk} J_k$ &\qquad  $ [J_i, K_j] = 
\epsilon_{ijk} K_k 
$ &\qquad$ [J_i, P_j] = \epsilon_{ijk} P_k  $ \\[0.2cm]
$[J_i, C_j] =  \epsilon_{ijk} C_k $ &\qquad  $[K_i, P_i] =\frac
1z({\rm e}^{zP_0}-1)  
$&\qquad  
$[K_i, C_i] =C_0-\frac z 2 \,\>K^2  $ \\[0.2cm]
$[P_0, D] = \frac 1z(1-{\rm e}^{zP_0})  $  & 
\qquad $[P_i, D] = -\,{\rm e}^{zP_0} P_i$&\qquad $ [P_0, C_i] =
-K_i$\\[0.2cm]
$[C_0, D] =C_0-\frac z2\,\>K^2  $&\qquad $[C_i, D] =  C_i + z
\,K_i D$ &\qquad $ [P_i, C_0] = K_i $\\[0.2cm]
\multicolumn{3}{l}{$ [P_i, C_j] = -\delta_{ij} D + 
\epsilon_{ijk}\,{\rm e}^{zP_0} J_k $\qquad\quad
$[C_i,C_j]=z(K_i C_j-K_jC_i) .$}
 \end{tabular}
\eeq
The Weyl--Carroll   subalgebra ${\cal {WC}}$,  $U_z({\cal
{WC}})$,  again remains as a Hopf subalgebra of $U_z({\cal
{C}})$.

On the other hand, although  divergencies within the
$\Phi$-factors (\ref{wg}) arise under the map (\ref{ma}),   both
the twisting element   (\ref{wi}) and the corresponding
universal ${\cal R}$-matrix   have   well defined limits given by
\beq\label{me}
\begin{array}{l}
{\cal F}=\exp(-  z K_3 \otimes   P_3)\exp(- z  K_2 \otimes   P_2 ) 
\exp(-z K_1\otimes P_1)\exp(-  z   D\otimes P_0) \\[2mm]
{\cal R}=\exp(-z P_3\otimes K_3)\exp(-z P_2\otimes K_2)\exp(-z
P_1\otimes K_1)\exp(- z P_0\otimes D)\cr
\qquad \qquad \times \exp(z D\otimes P_0)
\exp(z K_1\otimes P_1)\exp(z K_2\otimes  P_2)\exp(z K_3\otimes 
P_3). 
\label{dh}
\end{array}
\eeq

We stress that, at this dimension, the conformal Carroll
algebra is isomorphic to a $(4+1)$D Poincar\'e algebra
$iso(4,1)$; for a direct relationship a well adapted basis is
the $(4+1)$D version of the null-plane basis (\ref{goodness}).
In such a basis, $U_z({\cal C})$ turns out to be  the $(4+1)$D 
version of the $(3+1)$D null-plane quantum Poincar\'e algebra
obtained in  \cite{nulla,nullb}. We also recall that the
twisting element (\ref{dh}) for the $(3+1)$D case  was deduced
in \cite{Mudrov} by following a constructive method instead of a
contraction one.

\sect{Difference-differential massless Klein--Gordon equation}
\label{discretequation}

Non-standard quantum algebras have  already been 
shown to be Hopf symmetry algebras of several
difference-differential equations such as the $(1+1)$D
Schr\"odinger  \cite{herranz00},  $(1+1)$D wave and  2D
Laplace   \cite{herranz00a}, and $(3+1)$D wave and 4D Laplace
equations \cite{herranz02}.   In all these cases, discretization
arises in {\em one} (space or time) direction on a uniform
lattice, for which  the deformation parameter can be
interpreted   as the lattice  constant.  Our aim now is to show
that the generators of $U_z(so(4,2))$ can be expressed as 
discrete symmetries of  a  $(3+1)$D difference-differential
massless  Klein--Gordon (KG)  equation which, in turn, provides
a physical interpretation for the deformation parameter $z$ that
is slightly different with respect to the aforementioned known
cases.

Let us consider the Minkowkian spacetime with  coordinates
$x=(x^0,x^1,x^2,x^3)$ and   metric $(g_{\mu\nu})= {\rm
diag}\,(+,-,-,-)$ as introduced in section
\ref{conformalso42}.   As    is well known a differential
realization of its conformal Lie algebra
$so(4,2)$ with Lie brackets (\ref{so42}) is given by
\beq\begin{array}{l}  
\label{oa}
J_i =  \epsilon_{ijk} x^k \partial_j \qquad
       P_{\mu} = \partial_{\mu}  
\qquad K_i  =- x^i \partial_0 - x^0 \partial_i  \qquad
   D = -x^{\mu}\partial_{\mu} - \ell  \\[2mm]
C_0 = \frac{1}{2}\, x^\mu x_\mu \partial_0
-x^0(x^{\mu}\partial_{\mu}+
\ell)\qquad C_i = \frac{1}{2}\, x^\mu x_\mu \partial_i
+x^i(x^{\mu}\partial_{\mu}+
\ell)\end{array}
\eeq 
where $\partial_\mu={\partial}/{\partial x^\mu}$,
$x^\mu x_\mu=(x^0)^2-(x^1)^2-(x^2)^2-(x^3)^2$ and 
$\ell$ is called the conformal weight. Under this realization,
the Casimir operator of the Poincar\'e subalgebra, $ C = P_0^2 -
\>P^2$, provides the d'Alembertian $\square$ which leads to the
massless KG  equation:
\beq
\label{ob}
\square \phi(x) \equiv (\partial_0^2 - \partial_1^2 -
\partial_2^2  - \partial_3^2 ) \phi(x) = 0.
\eeq
  The following commutation relations between the d'Alembertian
and the operators (\ref{oa}), 
\beq \label{oc}
\begin{array}{ll}
[P_{\mu}, \square] = [J_i, \square] = [K_i, \square] = 0 
&\qquad  [D, \square] \ = 2\, \square  \\[2mm]
[C_{0}, \square] = 2 x^0\, \square + 2(\ell -1) \partial_0
&\qquad  [C_{i}, \square]\; =- 2x^i\, \square + 2(\ell -1) 
\partial_i ,
\end{array}\eeq
show that $so(4,2)$ is a Lie symmetry algebra of the massless
KG equation   whenever $\ell=1$. That is, its operators carry  
solutions of this equation into  solutions.
 
  Lie symmetries of the KG equation are deeply related with the
commutation relations $[\partial_{\mu}, x^{\nu}] =
\delta_{\mu\nu}$ (or  $[\partial_{\mu}, x_{\nu}] = g_{\mu\nu}$).
Hence, a procedure to obtain a difference-differential version
of (\ref{ob}) with $U_z(so(4,2))$-symmetry is first to deduce
some momentum $\opp_{\mu}=\opp_{\mu}(\partial,x)$ and position
$\opx^{\mu}=\opx^{\mu}(\partial,x)$ operators fulfilling
\beq
\label{oe}
[\opp_{\mu}, \opx^{\nu}] = \delta_{\mu\nu}  \qquad 
[\opp_{\mu}, \opx_{\nu}] = g_{\mu\nu}  \qquad  [\opx^{\mu},
\opx^{\nu}]  = 0 \qquad  [\opp_{\mu}, \opp_{\nu}] = 0 
\eeq
which in the limit $z\to 0$ should give $\opp^{\mu}\to
\partial_\mu$, $\opx^{\mu}\to x^\mu$, and second to express the
generators of $U_z(so(4,2))$ in terms of such operators.

The starting point in this construction comes from
the non-linear change of basis (\ref{wa})  applied in section
 \ref{twistdeformationso42conf} which naturally includes {\em
discrete derivatives}. Explicitly, if we write the generators
$\pp_\pm$, $\pp_a$ (\ref{ccarrier}) as the operators
$\opp^\mu$:
\beq
\label{ooff}
\begin{array}{l}
\pp_+=\frac 12(\opp_0+\opp_1)
\qquad
\pp_-=\frac 12(\opp_0-\opp_1)
\qquad
\pp_2=\frac 12(\opp_2-\opp_3)
\qquad
\pp_3=\frac 12(\opp_2+\opp_3)   
\end{array}
\eeq
and  consider the translations generators $P_\mu$ introduced by
means of the non-linear map (\ref{wa}) as the derivatives
$\partial_\mu$,  then the transformed expressions for $\pp_\pm$,
$\pp_a$ given in (\ref{wb})  allow  us to make the  following
ansatz  for the $\opp_{\mu}$ operators:
\beq\label{of}
\begin{array}{l}
\opp_0+\opp_1= \frac 1z({1-  {\rm e}^{- z\partial_0}\,{\rm e}^{
-z\partial_1}})-z \,{\rm e}^{-z\partial_0}\,{\rm
e}^{z\partial_1}(\partial_2^2+\partial_3^2)\\[0.2cm]
\opp_0-\opp_1=\frac 1z(1-  {\rm e}^{- z\partial_0}\,{\rm e}^{
z\partial_1}) \\[0.2cm]
\opp_2\pm \opp_3= {\rm e}^{-z\partial_0}\,{\rm
e}^{z\partial_1} (\partial_2 \pm\partial_3)  .
\end{array}
\eeq
In other words,  $\sigma_\pm=-z(\partial_0\pm\partial_1)$ and
the elements $\pp_-$ and $\pp$ are directly two discrete
derivatives along two basic directions $x^0-x^1$ and $x^0+x^1$,
respectively.

Next, by requiring commutators (\ref{oe}) to be satisfied, we
find the corresponding $\opx^\mu$ operators:
\beq
\label{og}
\begin{array}{l}
 \opx^{0} + \opx^{1} = (x^0+x^1) {\rm e}^{z\partial_0}\,{\rm
e}^{z\partial_1}\\[0.2cm]
\opx^{0} - \opx^{1} = \left\{ x^0-x^1 + 2 z (x^2\partial_2 +
x^3\partial_3 ) \right\} {\rm e}^{z\partial_0}\,{\rm
e}^{-z\partial_1} +z^2(x^0+x^1) {\rm e}^{z\partial_0}\,{\rm
e}^{z\partial_1}(\partial_2^2+\partial_3^2) \\[0.2cm]
\opx^2\pm   \opx^3= (x^2\pm x^3){\rm e}^{z\partial_0}\,{\rm
e}^{-z\partial_1} +z (x^0+x^1){\rm e}^{z\partial_0}\,{\rm
e}^{z\partial_1}(\partial_2\pm
\partial_3) .
\end{array}
\eeq
 Therefore, from (\ref{of}) and (\ref{og}) we find that:
\beq
\label{oh}
\begin{array}{l}
\opp_0=\frac 1z(1-{\rm e}^{-z\partial_0}\cosh
z\partial_1)-\frac z2\,{\rm e}^{-z\partial_0}\,{\rm
e}^{z\partial_1} (\partial_2^2+\partial_3^2)\\[0.2cm]
\opp_1=\frac 1z\,{\rm e}^{-z\partial_0}\sinh z\partial_1 -\frac
z2\,{\rm e}^{-z\partial_0}\,{\rm e}^{z\partial_1}
(\partial_2^2+\partial_3^2)  \\[0.2cm]  
\opp_a = {\rm e}^{-z\partial_0}\,{\rm
e}^{z\partial_1}  \partial_a\\[0.2cm]
 \opx^{0}  = x^0\,{\rm e}^{z\partial_0}\cosh z\partial_1 +
x^1\,{\rm e}^{z\partial_0}\sinh z\partial_1+z (x^2\partial_2+
x^3\partial_3) {\rm e}^{z\partial_0}\,{\rm
e}^{-z\partial_1}\\[0.1cm]
\qquad\qquad +\frac {z^2}2(x^0+x^1)
{\rm e}^{z\partial_0}\,{\rm
e}^{z\partial_1}(\partial_2^2+\partial_3^2)\\[0.2cm]
 \opx^1  = x^1\,{\rm e}^{z\partial_0}\cosh z\partial_1 +
x^0\,{\rm e}^{z\partial_0}\sinh z\partial_1-z (x^2\partial_2+
x^3\partial_3) {\rm e}^{z\partial_0}\,{\rm
e}^{-z\partial_1}\\[0.1cm]
\qquad\qquad -\frac {z^2}2(x^0+x^1)
{\rm e}^{z\partial_0}\,{\rm
e}^{z\partial_1}(\partial_2^2+\partial_3^2)\\[0.2cm]
\opx^a =  x^a\,{\rm e}^{z\partial_0}\,{\rm e}^{-z\partial_1}
+z (x^0+x^1){\rm e}^{z\partial_0}\,{\rm e}^{z\partial_1}
\partial_a   
\end{array}
\eeq
where $a=2,3$. If we now apply the maps $\partial_\mu \mapsto 
\opp_\mu$, $x^\mu\mapsto  \opx^\mu$ to the initial differential
operators (\ref{oa}) and KG equation  (\ref{ob}), we obtain
difference-differential symmetries 
$X(\partial_\mu,x^\mu)\mapsto {\hat X}(\opp_\mu,\opx^\mu)$ of 
the following difference-differential  analogue of the 
massless KG equation:
\beq
  \square_z \phi(x) = \left\{
\left( \frac{ 1-{\rm e}^{-z\partial_0}\,{\rm
e}^{-z\partial_1}}{z}\right)
\left( \frac{ 1-{\rm e}^{-z\partial_0}\,{\rm
e}^{z\partial_1}}{z}\right)
  - {\rm e}^{-z\partial_0}\,{\rm e}^{z\partial_1} (\partial_2^2
+ \partial_3^2) \right\} \phi(x) = 0 
  \label{oi}
\eeq
which  is the generalization of the $(2+1)$D  case studied in
\cite{aizawa01b}. The transformed commutators (\ref{oc}) with
$\ell=1$  show that $U_z(so(4,2))$ is a  symmetry twisted
algebra of this equation. Coproduct for the resulting operators
can be deduced from the expressions given in section
\ref{twistdeformationso42}  through the change of basis
(\ref{ccarrier})--(\ref{restoc}). This can be further used in
the construction of  composition of several
difference-differential  massless KG equations.

On the other hand, relations (\ref{of}) and (\ref{og}) indicate
that an alternative and natural basis to rewrite the above
results is the Minkowskian null-plane one (\ref{nullcoo}). Then
by denoting
\beq
\!\!\!\begin{array}{l}
\partial_\pm=\partial_0\pm \partial_1\qquad
[\partial_\pm,x^\pm]=1
\qquad [\partial_\pm,x^\mp]=0\qquad
\opp_\pm=\opp_0\pm \opp_1\qquad
\opx^\pm=\frac 12(\opx^0\pm \opx^1)\qquad
\end{array}
  \label{oj}
\eeq
we find the following null-plane momentum and position operators 
\beq
\label{ok}
\begin{array}{l}
\opp_+= \frac 1z(1-  {\rm e}^{- z\partial_+})-z \,{\rm
e}^{-z\partial_-} (\partial_2^2+\partial_3^2)\qquad\quad \opx^+
= x^+\,{\rm e}^{z\partial_+}\\[0.2cm]
\opp_-=\frac 1z(1-  {\rm e}^{- z\partial_-} ) \qquad\ 
\opx^- =  x^- {\rm e}^{z\partial_-}+ z (x^2\partial_2 +
x^3\partial_3 ) {\rm e}^{z\partial_-} +z^2 x^+ {\rm
e}^{z\partial_+}(\partial_2^2+\partial_3^2) \\[0.2cm]
\opp_a= {\rm e}^{-z\partial_-} \partial_a 
\qquad\quad\quad\;\;
\opx^a =  x^a\,{\rm e}^{z\partial_-}
+ 2 z  x^+ {\rm e}^{z\partial_+} \partial_a    .
\end{array}
\eeq
 that verify  (\ref{oe}) for $\mu,\nu=+,-,2,3$. In this
null-plane context the  massless KG equation (\ref{oi}) 
is rewritten as
\beq
  \square_z \phi(x) = \left\{
  \left( \frac{ 1-{\rm e}^{-z\partial_+}}{z}\right)
\left( \frac{ 1-{\rm e}^{-z\partial_-}}{z}\right)
   - {\rm e}^{-z\partial_-} (\partial_2^2 + \partial_3^2)
  \right\} \phi(x^+,x^-,x^2,x^3) = 0 
  \label{ol}
\eeq
where the action of a discrete derivative reads
\beq
\begin{array}{l}
 \frac 1z ({ 1-{\rm e}^{-z\partial_+}})\phi(x^+,x^-,x^2,x^3)
=\frac 1z \left\{ \phi(x^+,x^-,x^2,x^3)
-\phi(x^+ - z,x^-,x^2,x^3)\right\}
\end{array} 
\label{om}
\eeq
and similar for that with $\partial_-$. Consequently, we 
clearly obtain an intrinsic discretization on a uniform lattice
along   the {\em two} null-plane directions $x^\pm$ where $z$ is
the {\em same} lattice step for both of them; the transverse
coordinates $x^a$ remain as continuous ones. Furthermore, the
null-plane evolution framework with $x^-=\tau$ (\ref{nullcoo})
also provides an interpretation of $z$ as a fundamental constant
directly associated to a discretization of the `time' evolution
parameter $\tau$ so that  the transformed operator
$\hat{P}_-(\opp_\mu,\opx^\mu)$ would play the role of a
difference-differential Hamiltonian \cite{nulla}.

\sect{Conclusions and remarks}

 In this paper we have constructed a new quantum twisted
deformation of the conformal algebra of the $(3+1)$D Minkowskian
spacetime that generalizes the previously known lower
dimensional cases \cite{beyond,herranz97}. The twisting element
has been established within a conformal null-plane basis. 
Moreover, expressions in the standard conformal basis have been
deduced through a non-linear map and, among them, we remark the
`compact' form of the universal ${\cal R}$-matrix of
$U_z(so(4,2))$. 

These results have allowed us to perform a study of contractions
that relate in a consistent way $U_z(so(4,2))$ with its
non-relativistic limits together with their  Weyl Hopf
subalgebras, classical $r$-matrices, twisting elements and 
universal ${\cal R}$-matrices. In particular, the twisted
conformal Carroll algebra is   the $(4+1)$D version of the so
called null-plane quantum Poincar\'e algebra
\cite{nulla,nullb}. 

As an application, it has been also shown that $U_z(so(4,2))$ is
the symmetry Hopf algebra of a  difference-differential
 massless KG equation, for which, the symmetry operators have
been deduced; this has provided an interpretation of the
deformation parameter $z$ as the lattice constant in a
discretization of the $(3+1)$D Minkwoskian spacetime along {\em
two} null-plane directions. 

All of these results are summarized in the following diagram,
where vertical arrows indicate  contractions.
$$
\begin{array}{ccl}
\mbox{Twisted conformal Galilean algebra}&\quad U_z({\cal
{WG}})  &\subset\   U_z({\cal {G}})\\[2mm]
&\qquad \uparrow&\qquad\quad\uparrow\\[2mm]
\mbox{Twisted conformal Minkowskian algebra}&\quad U_z({\cal {WP}}) 
&\subset\   U_z(so(4,2)) \leftrightarrow \square_z \phi(x)=0 \\[2mm]
&\qquad \downarrow&\qquad\quad\downarrow\\[2mm]
\mbox{Twisted conf.\ Carroll/Null-plane Poincar\'e}&\quad
U_z({\cal {WC}})  &\subset\   U_z({\cal {C}})\simeq U_z(iso(4,1))
 \end{array} 
$$

To end with we would like to comment on  a twisting deformation
of $sl(4)$   presented  by   Kulish and Lyakhovsky (KL)
in~\cite{kulish00} as well as on  the classical $r$-matrices for
$so(4,2)$ obtained by Lukierski,  Lyakhovsky and Mozrzymas (LLM) in
\cite{Lukierskid}

The KL deformation is  also based  on an 8D carrier
 subalgebra $\>L$ of $sl(4)$,  which as a {\em complex} algebra is 
isomorphic to $so(6)$ (although we have dealt with  the real
form $so(4,2)$).   When   both carrier algebras, $\>L$ and $\cal L$
(\ref{carrier}), are considered over the complex numbers,  there
is  an   isomorphism  $\>L \simeq \cal L$ defined by the following
change of basis:
\beq
\begin{array}{ll}
H_{23}=H\equiv \hh_+ &\qquad
H_{14}=H_{\lambda_0^{\bot}}\equiv \hh_- \\[0.2cm]
E_{23}= L_{\lambda_0}\equiv -\frac{1}{2 z}\,\pp_+&\qquad
E_{14}= L_{\lambda_0^{\bot}}\equiv -{2 z}\, \pp_-\\[0.2cm]
E_{13}=  L_{\lambda_0^{\bot}-\tilde \lambda'}   \equiv \frac
1{\sqrt{2}}\, (\pp_3- i\pp_2)&\qquad
E_{24}= L_{\lambda_0^{\bot}-\lambda'} \equiv \frac 1{\sqrt{2}}\,
(\pp_3+ i\pp_2)\\[0.2cm]
E_{12}= L_{\lambda'}  \equiv - z {\sqrt{2}}\,
(\bb_3- i\bb_2)&\qquad
E_{34}=- L_{\tilde \lambda'} \equiv  z{\sqrt{2}}\,
(\bb_3+ i\bb_2) 
\end{array} 
\label{on}
\eeq
where the KL generators are written on the l.h.s. in two
notations; starting from the commutation relations of
$\>L$ we recover those of $\cal L$ (\ref{carrier}).
If now we apply  this change of basis to the KL
twisting element, here written as
\beq
{\cal F}_{\rm KL}=\Psi_{{\cal BJ}}\Psi_{{\cal E'}}\Psi_{{\cal
E}}\Psi_{{\cal J}_\bot}
\label{op}
\eeq
we also find that this coincides exactly with   ${\cal F}$ 
(\ref{twist}), for which the $\Phi$-factors (\ref{twistfactor})
are related with the KL
$\Psi$-factors by means of
\beq
\Psi_{{\cal J}_\bot}\equiv \Phi_{{\cal J}}\qquad
\Psi_{{\cal E'}}\Psi_{{\cal E}}\equiv  \Phi_{{\cal E}}\qquad
\Psi_{{\cal BJ}}\equiv  \Phi_{{\cal A}} 
\label{oq}
\eeq
provided that
\beq
\sigma_{14}\equiv \sigma_-\qquad
\sigma_{{\cal B}}\equiv \sigma_+\qquad
E_{\cal B}\equiv  2 z\pp .
\label{or}
\eeq
Consequently, the twisting   $\>L$  and ${\cal L}$   algebras
are   identified.

The LLM  deformations are  provided by classical $r$-matrices also
defined on  carrier subalgebras   of $sl(4)\simeq so(6)$ which give
rise to  $\kappa$-deformations of ${\cal {P}}$, ${\cal {WP}}$ and
$so(4,2)$ in a conformal basis  by imposing reality conditions. The
so called light-like $\kappa$-Poincar\'e
$r$-matrix is expressed in our basis (\ref{so42}) as
\beq
 r_{LLM}=z K_3\wedge P_0  +z( K_1\wedge P_1 + K_2\wedge P_2 +
K_3\wedge P_3)+z(J_1\wedge P_2 - J_2 \wedge P_1)
\label{Lukierresa}
\eeq
where $z=1/\kappa$. By comparing $r_{LLM}$ with our $r$
(\ref{ac}), we find that the former  is associated with the boost
generator $K_3$ instead of $K_1$, but  the relevant difference
among them lies in the replacement $K_3\wedge P_0\leftrightarrow
D\wedge P_0$.   In fact, $r$ (\ref{ac})  may be seen as a
light-like $\kappa$-Weyl--Poincar\'e $r$-matrix.  Another more
involved LLM classical $r$-matrix is the $\kappa$-Weyl
 one that reads  
\beq
\begin{array}{l}
\hat r_{LLM}=r_{LLM}+\frac z2 (D-K_3)\wedge (P_0+P_3) \\[2mm]
 \qquad\ \ =\frac z2 K_3\wedge P_0 +\frac z2 D\wedge (P_0+P_3)\\[2mm]
\qquad \qquad \ 
 +z( K_1\wedge
P_1 + K_2\wedge P_2 +
\frac 12 K_3\wedge P_3)+z(J_1\wedge P_2 - J_2 \wedge P_1) .
\end{array}
\label{Lukierresb}
\eeq
 Hence the corresponding LLM-twisted deformations determined by
(\ref{Lukierresa}) and (\ref{Lukierresb}) would be different from
the one here studied.

\section*{Acknowledgments}

This work has been partially supported by
  the Ministerio de Ciencia y Tecnolog\'{\i}a of Spain under
 Projects PB98-0360 and  BFM2000-1055. N.A. was supported  by the
Ministry of Education, Culture, Sports, and Technology, Japan.


\end{document}